\documentclass[12pt]{article}
\usepackage{epsfig}
\usepackage{multirow}

\newcommand{\mysection}{\setcounter{equation}{0}\section}

\def\beq{\begin{equation}}
\def\eeq{\end{equation}}
\def\beqa{\begin{eqnarray}}
\def\eeqa{\end{eqnarray}}
 
\newlength{\dinwidth} \newlength{\dinmargin}
\begin{document}

\begin{center}
{\Large \bf NNNLO soft-gluon corrections for the top-antitop pair production cross section}
\end{center}
\vspace{2mm}
\begin{center}
{\large Nikolaos Kidonakis}\\
\vspace{2mm}
{\it Kennesaw State University,  Physics \#1202,\\
Kennesaw, GA 30144, USA}
\end{center}
 
\begin{abstract}
I present a calculation of next-to-next-to-next-to-leading-order (NNNLO) 
soft-gluon corrections for top-antitop pair production in hadronic collisions. 
Approximate NNNLO (aNNNLO) results are obtained by adding the NNNLO soft-gluon corrections to the complete next-to-next-to-leading-order (NNLO) cross section. 
Theoretical predictions are shown for the total aNNNLO $t{\bar t}$ 
cross section at LHC and Tevatron energies. The aNNNLO cross sections are 
larger but have smaller theoretical uncertainties than at NNLO.
\end{abstract}
 
\mysection{Introduction}

The top quark retains a leading position among elementary particles as the heaviest particle that has been discovered so far. Top-quark physics occupies a prominent role in the LHC program as well as past experiments at the Tevatron. Top quark studies, both in top-antitop pair production and in single-top production, are important for QCD and electroweak physics, and many measurements of total cross sections, differential distributions, and top quark properties have been performed at the Tevatron and the LHC. The large mass of the top quark suggests that it plays an important role in Higgs physics. The top quark also figures prominently in searches for flavor-changing neutral currents and in searches for other physics beyond the Standard Model. 

The experimental measurements of the top quark total cross section, transverse momentum ($p_T$) distributions, and rapidity distributions,  
are currently in good agreement with theoretical predictions. 
The experimental errors continue to decrease with time, which necessitates 
increasingly precise theoretical calculations. 
Next-to-leading order (NLO) calculations of the $t{\bar t}$ cross section 
have been available for a long time \cite{NLO1,NLO2,NLO3}. 
The NLO predictions have been improved by    
the inclusion of higher-order soft-gluon corrections through resummations, first at next-to-leading-logarithm (NLL) accuracy and, more recently, at next-to-next-to-leading-logarithm (NNLL) accuracy. A brief history of these soft-gluon threshold resummations and a large list of references can be found in the review paper of Ref. \cite{NKBP}. Approximate NNLO results have also been obtained from those resummations. In addition a large effort has been invested in the calculation of complete NNLO corrections (see again \cite{NKBP} for more details and references). At present there is still not a complete analytical calculation at NNLO, but numerical results for the total $t{\bar t}$ cross section at NNLO appeared in \cite{CFM}. 

There are many differences between various resummation approaches 
in the literature and these have been detailed previously in \cite{NKBP}. 
Some resummations are for double-differential cross sections while others are only 
for total cross sections; some resummation formalisms use moment-space perturbative 
QCD (pQCD) while others use Soft-Collinear Effective Theory (SCET). There are also 
additional technical differences among the various methods \cite{NKBP}.

The double-differential resummation approach uses a more general definition of threshold and it allows the calculation of transverse momentum and rapidity distributions, in addition to total cross-section calculations. In Ref. \cite{NK10} NNLL resummation for the double-differential cross section using  moment-space pQCD was used to 
derive approximate NNLO results for the total $t{\bar t}$ cross section and top-quark differential 
distributions. Now that exact NNLO numerical results for the total cross section are available, one can improve
the theoretical prediction for that quantity by adding aNNNLO corrections from 
soft-gluon resummation.  

The validity, stability, and reliability of the theoretical predictions from our resummation formalism  
has been discussed several times before, most recently in Ref. \cite{NKHQ13}.
The NLO and NNLO soft-gluon corrections from our formalism in Refs. \cite{NK10,NKHQ13} are very close to the exact results.
The numerical accuracy of the threshold soft-gluon approximation was first demonstrated by comparing exact NLO and approximate NLO results  
for the total $t{\bar t}$ cross section and the top quark $p_T$ and rapidity distributions. 
If one compares the total NLO cross section 
the difference between approximate and exact results is entirely negligible, 
well below 1\%. This excellent agreement between exact and approximate NLO results 
is true not only for the total cross section but also for the differential distributions 
(see Fig. 11 and related discussion in \cite{NKHQ13}).

We can also make comparisons at NNLO for the total $t{\bar t}$ cross section. Our approximate NNLO cross sections \cite{NK10,NKHQ13} are very close to the exact NNLO \cite{CFM} results: both the central values and the scale uncertainty are nearly 
the same and this holds true for all LHC and Tevatron collider energies and 
top quark masses (see Figs. 13 and 14 and the corresponding discussion in \cite{NKHQ13}).
There is less than 1\% difference between approximate and exact cross sections at 
both NLO and NNLO, and our results approximate the exact results better than other methods such as the one used in \cite{CFM}. 
This is a very important fact since it displays the validity and relevance of our method and it provides strong 
confidence that the aNNNLO soft-gluon corrections will be good approximations 
to the exact results at that higher order. Also the comparison of our results 
with Tevatron and LHC total cross sections and $p_T$ and rapidity distributions shows excellent agreement between our theoretical predictions and the experimental data (see e.g. Ref. \cite{NKHQ13}).

In the next section we give some more details of our calculation and present analytical results
for the aNNNLO corrections. Section 3 includes all our numerical results for the $t{\bar t}$ total
cross section at 1.96 TeV Tevatron energy and at 7, 8, 13, and 14 LHC energies. 
Our aNNNLO predictions are found by adding the NNNLO soft-gluon corrections to the complete NNLO $t{\bar t}$ total cross section.
We conclude in Section 4.

\mysection{Analytical results}

Soft-gluon resummation follows from the renormalization group evolution 
of functions describing  collinear and soft-gluon emission
in the factorized partonic cross section, and it involves the details of the 
color structure of the process. The resummed cross section can be expanded at 
fixed order in the strong coupling, $\alpha_s$, to NLO, NNLO, NNNLO, etc.,  
and inverted back to momentum space. The necessary ingredients for our 
calculation can be found in Refs. \cite{NK10,NKHQ13,NKNNNLO}.

At leading order (LO) the two partonic channels for $t{\bar t}$ production are 
quark-antiquark annihilation
\beq
q(p_1)+{\bar q}(p_2) \rightarrow t(p_3) +{\bar t}(p_4)
\eeq
and gluon-gluon fusion
\beq
g(p_1)+g(p_2) \rightarrow t(p_3) +{\bar t}(p_4)
\eeq

We define the kinematical variables $s=(p_1+p_2)^2$,  $t_1=(p_1-p_3)^2-m_t^2$, 
$u_1=(p_2-p_3)^2-m_t^2$, and $\beta=\sqrt{1-4m_t^2/s}$, where $\beta$ is the 
top-quark speed in LO kinematics and $m_t$ is the top-quark mass. 
We also define the threshold variable $s_4=s+t_1+u_1$. At partonic threshold 
there is no energy available for additional radiation and $s_4$ vanishes in that limit. It is 
important to note that partonic threshold is a more general concept than absolute threshold, 
the latter meaning that the top quark is produced at rest. 
In other words if $s_4=0$ that does not mean that $\beta$ has to be zero; 
partonic threshold can be reached even with arbitrarily high transverse momentum of the top quark.
In our resummation the threshold limit is always given by $s_4 \rightarrow 0$. 

At each order in $\alpha_s$, one encounters plus-distribution terms of the form 
$[\ln^k(s_4/m_t^2)/s_4]_+$ and, for the $n$-th order corrections, 
the power of the logarithm, $k$, can range from the leading value of $2n-1$  down to the lowest value of 0. 
Thus, at NLO the leading value of $k$ is 1, at NNLO it is 3, 
and at NNNLO it is 5.
The expressions for the NNNLO soft-gluon corrections fully written out are 
extremely long, although they can be written in a more compact implicit form as
first shown in \cite{NKNNNLO}. However, the first two powers of the logarithms 
have coefficients that are relatively short and we display them below. 

For $q {\bar q} \rightarrow t{\bar t}$, the NNNLO soft-gluon corrections 
to the double-differential cross section are 
\beqa
\frac{d^2\sigma^{(3)}_{q {\bar q}\rightarrow t{\bar t}}}{dt_1 \, du_1}&=&
\frac{\alpha_s^5}{\pi^2} \frac{C_F}{N_c \, s^2} \left(\frac{t_1^2+u_1^2}{s^2}
+\frac{2m_t^2}{s}\right)
\left\{8 C_F^3 \left[\frac{\ln^5(s_4/m_t^2)}{s_4}\right]_+ \right.
\nonumber \\ && 
{}+\left[20 C_F^3 \left(4\ln\left(\frac{t_1}{u_1}\right)
-\ln\left(\frac{t_1u_1}{m_t^4}\right)-\ln\left(\frac{\mu_F^2}{s}\right)
-\frac{(1+\beta^2)}{2\beta}\ln\left(\frac{1-\beta}{1+\beta}\right) -1\right)
\right.
\nonumber \\ && \quad
{}+10C_F^2 C_A \left(-3 \ln\left(\frac{t_1}{u_1}\right)
+\ln\left(\frac{t_1u_1}{s m_t^2}\right)+\frac{(1+\beta^2)}{2\beta}
\ln\left(\frac{1-\beta}{1+\beta}\right)-\frac{11}{9}\right)
\nonumber \\ && \quad \left. \left.
{}+\frac{20}{9} C_F^2 n_f\right]
\left[\frac{\ln^4(s_4/m_t^2)}{s_4}\right]_+  \right\}
\nonumber \\ && 
{}+ \cdots 
\eeqa
where $C_F=(N_c^2-1)/(2N_c)$, with $N_c=3$ the number of colors, $C_A=N_c$, 
and $n_f=5$ is the number of light-quark flavors. As discussed above, 
we do not display cubic or lower logarithmic powers, which have very long 
expressions for their coefficients.

For $gg \rightarrow t{\bar t}$, the NNNLO soft-gluon corrections 
to the double-differential cross section are
\beqa
\frac{d^2\sigma^{(3)}_{gg\rightarrow t{\bar t}}}{dt_1 \, du_1}&=&
\frac{\alpha_s^5}{\pi^2} \frac{2N_c \, C_F}{(N_c^2-1)^2 \, s^2} 
\left(C_F-C_A \frac{t_1 \, u_1}{s^2}\right)
\left[\frac{t_1}{u_1}+\frac{u_1}{t_1}+\frac{4m_t^2 \,s}{t_1 \, u_1} 
\left(1-\frac{m_t^2 \, s}{t_1 \, u_1}\right)\right]
\nonumber \\ && \hspace{-5mm} \times
\left\{8 C_A^3 \left[\frac{\ln^5(s_4/m_t^2)}{s_4}\right]_+ \right.
\nonumber \\ &&  \hspace{-5mm} \quad \left.
{}+\left[-20 C_A^3 \left(\ln\left(\frac{t_1u_1}{m_t^4}\right)
+\frac{11}{18}\right)
-20\, C_A^2 C_F \ln\left(\frac{\mu_F^2}{s}\right)
+\frac{20}{9} C_A^2 n_f\right]
\left[\frac{\ln^4(s_4/m_t^2)}{s_4}\right]_+ \right\}
\nonumber \\ && \hspace{-10mm}
{}+\frac{\alpha_s^5}{\pi^2} \frac{10 C_A^2}{(N_c^2-1) \, s^2} 
\left[\frac{t_1}{u_1}+\frac{u_1}{t_1}+\frac{4m_t^2s}{t_1u_1}
\left(1-\frac{m_t^2s}{t_1u_1}\right)\right] 
\nonumber \\ && \hspace{-5mm} \times
\left\{N_c \left(1-\frac{2t_1u_1}{s^2}\right)
\left[\left(-C_F+\frac{C_A}{2}\right) \frac{(1+\beta^2)}{2\beta}
\ln\left(\frac{1-\beta}{1+\beta}\right) -C_F
+\frac{N_c}{2} \ln\left(\frac{t_1u_1}{m_t^2 s}\right)\right] \right.
\nonumber \\ && \hspace{-5mm} \quad 
{}+\frac{1}{N_c}(C_F-C_A) \frac{(1+\beta^2)}{2\beta}
\ln\left(\frac{1-\beta}{1+\beta}\right)+\frac{C_F}{C_A}
-\ln\left(\frac{t_1u_1}{m_t^2 s}\right)
\nonumber \\ && \hspace{-5mm} \quad \left.
{}+\frac{N_c^2}{2} \frac{(t_1^2-u_1^2)}{s^2} \ln\left(\frac{u_1}{t_1}\right)
\right\} \left[\frac{\ln^4(s_4/m_t^2)}{s_4}\right]_+ 
\nonumber \\ && \hspace{-10mm}
{}+ \cdots
\eeqa
where again we have only explicitly shown the first two powers of the 
logarithms.

We employ the above analytical results to calculate the aNNNLO $t{\bar t}$  
cross sections at the LHC and the Tevatron in the next section.

\mysection{Total $t{\bar t}$ cross sections at the LHC and the Tevatron}

We now provide a study of the top-antitop pair production cross sections  
at the Tevatron and LHC colliders.
We present approximate NNNLO (aNNNLO) calculations for these quantities. 
The aNNNLO results are computed by adding the NNNLO soft-gluon 
corrections (derived from NNLL resummation) to the exact NNLO quantities.
We use the MSTW2008 NNLO parton distibution functions (pdf) \cite{MSTW2008}.

\begin{table}[htb]
\begin{center}
\begin{tabular}{|c|c|c|c|c|c|} \hline
\multicolumn{6}{|c|}{aNNNLO  $t{\bar t}$ cross section (pb)} \\ \hline
$m_t$ (GeV) & Tevatron & LHC 7 TeV & LHC 8 TeV & LHC 13 TeV& LHC 14 TeV \\ \hline
170 & 8.18 & 192 & 273 & 886 & 1046 \\ \hline 
171 & 7.92 & 186 & 265 & 862 & 1018 \\ \hline 
172 & 7.67 & 181 & 257 & 839 & 991 \\ \hline 
173 & 7.44 & 175 & 250 & 816 & 964 \\ \hline 
174 & 7.20 & 170 & 243 & 794 & 939 \\ \hline 
175 & 6.98 & 165 & 236 & 773 & 914 \\ \hline 
\end{tabular}
\caption[]{The aNNNLO $t{\bar t}$ production cross section in pb 
in $p{\bar p}$ collisions at the Tevatron with $\sqrt{S}=1.96$ TeV and   
in $pp$ collisions at the LHC with $\sqrt{S}=7$, 8, 13, and 14 TeV. We set 
$\mu=m_t$ and use the MSTW2008 NNLO pdf \cite{MSTW2008}.}
\label{table1}
\end{center}
\end{table}

Table \ref{table1} lists the central values for the aNNNLO $t{\bar t}$
cross sections at Tevatron and LHC energies for top quark masses 
between 170 GeV and 175 GeV. 
For these central values we set the factorization and renormalization scales
$\mu=m_t$ and we use the central pdf set of \cite{MSTW2008}.
There are two kinds of theoretical uncertainties associated with the 
calculation: dependence on the factorization and renormalization scales, 
and uncertainties from the parton distributions. 
We provide more results and discuss these uncertainties in more detail 
for each collider energy below.

\begin{figure}
\begin{center}
\includegraphics[width=11cm]{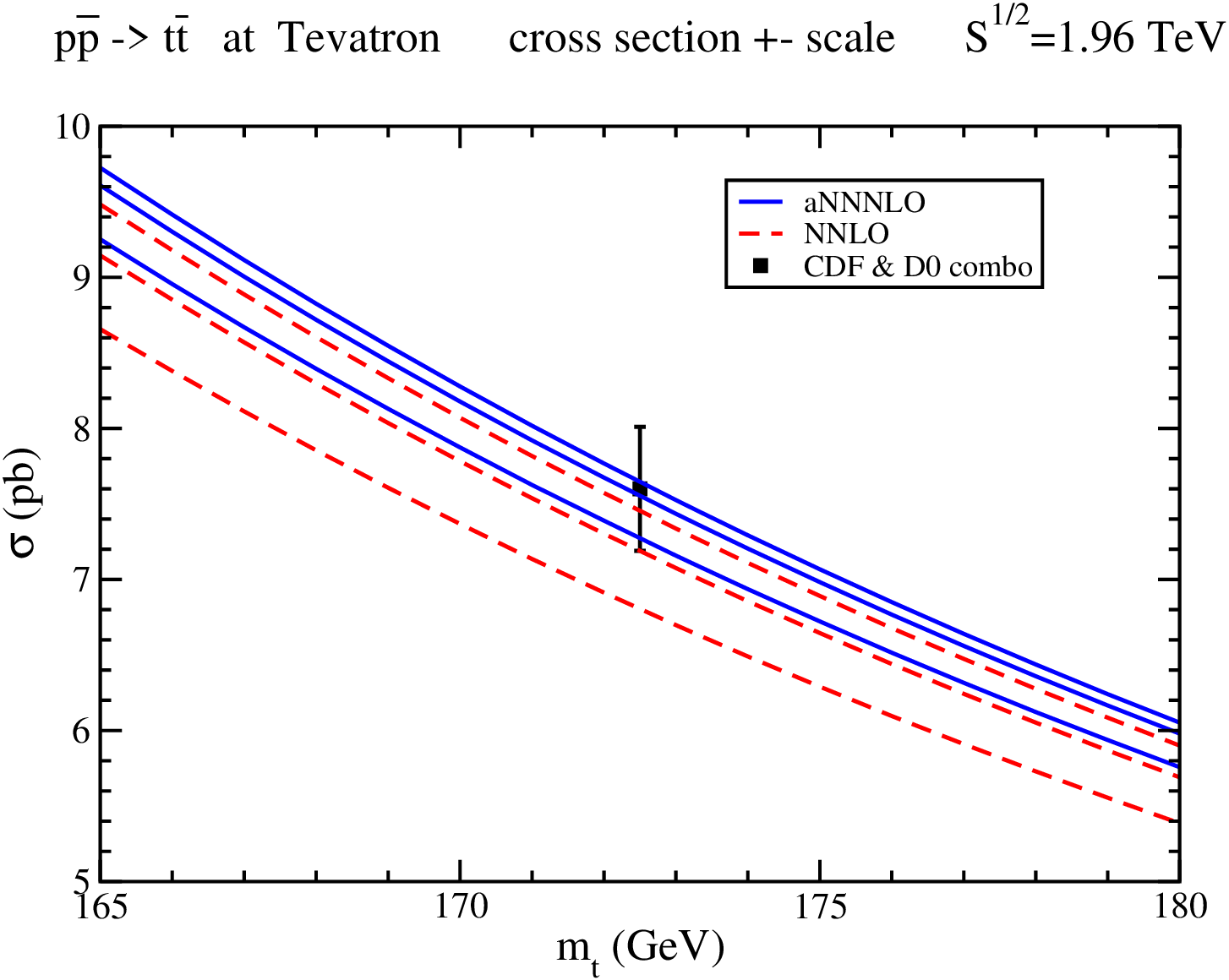}
\caption{The NNLO and aNNNLO total cross sections for $t{\bar t}$ production 
at the Tevatron with $\sqrt{S}=1.96$ TeV and comparison to CDF and D0 
combination data \cite{Tevatron}. The central lines at each order are with 
$\mu=m_t$, and the other lines display the upper and lower values from scale 
variation.}
\label{toptev}
\end{center}
\end{figure}

We begin with a study of top-antitop production in proton-antiproton 
collisions at the Tevatron at 1.96 TeV energy.
In Fig. \ref{toptev} we plot the cross section over a top quark mass 
range $165 \le m_t \le 180 $ GeV at NNLO and aNNNLO. The central result at 
each order is with a factorization and renormalization scale $\mu=m_t$. 
The independent variation with scales over the interval from $m_t/2$ to $2m_t$ 
is shown by the upper and lower lines at each order.  
It is evident that the theoretical scale dependence is smaller at aNNNLO than 
at NNLO, as expected. The comparison with a recent combination of CDF and D0 
Tevatron data \cite{Tevatron} shows a nice agreement between theory and 
experiment. The enhancement from the aNNNLO corrections is significant and 
improves the description of the Tevatron data.

For a top quark mass of 173.3 GeV, the aNNNLO cross section at the Tevatron is
\beq
\sigma^{\rm aNNNLO}_{p{\bar p} \rightarrow t{\bar t}}(m_t=173.3 \, {\rm GeV}, \, \sqrt{S}=1.96\, {\rm TeV})
=7.37 {}^{+0.09}_{-0.27} {}^{+0.38}_{-0.28}=7.37 \pm 0.39 \; {\rm pb} \, .
\eeq
Here the first uncertainty is from scale variation over $0.5 \le \mu/m_t \le 2$ 
and the second is from the MSTW2008 NNLO pdf errors at 90\% C.L. (to be 
conservative, we do not use the smaller 68\% C.L. pdf errors). 
The total uncertainty number is found by adding the scale and pdf errors in 
quadrature, and it is $\pm$5.3\% at the Tevatron. We also find that the aNNNLO
result provides an enhancement of 5.1\% over the NNLO value. Thus both the 
overall increase of the cross section and the decrease in its uncertainty at 
aNNNLO relative to NNLO are significant.

We continue with a study of top quark 
production in proton-proton collisions at the LHC. We present results for 
the  past LHC energies of 7 and 8 TeV 
and the future LHC energies of 13 and 14 TeV.

\begin{figure}
\begin{center}
\includegraphics[width=11cm]{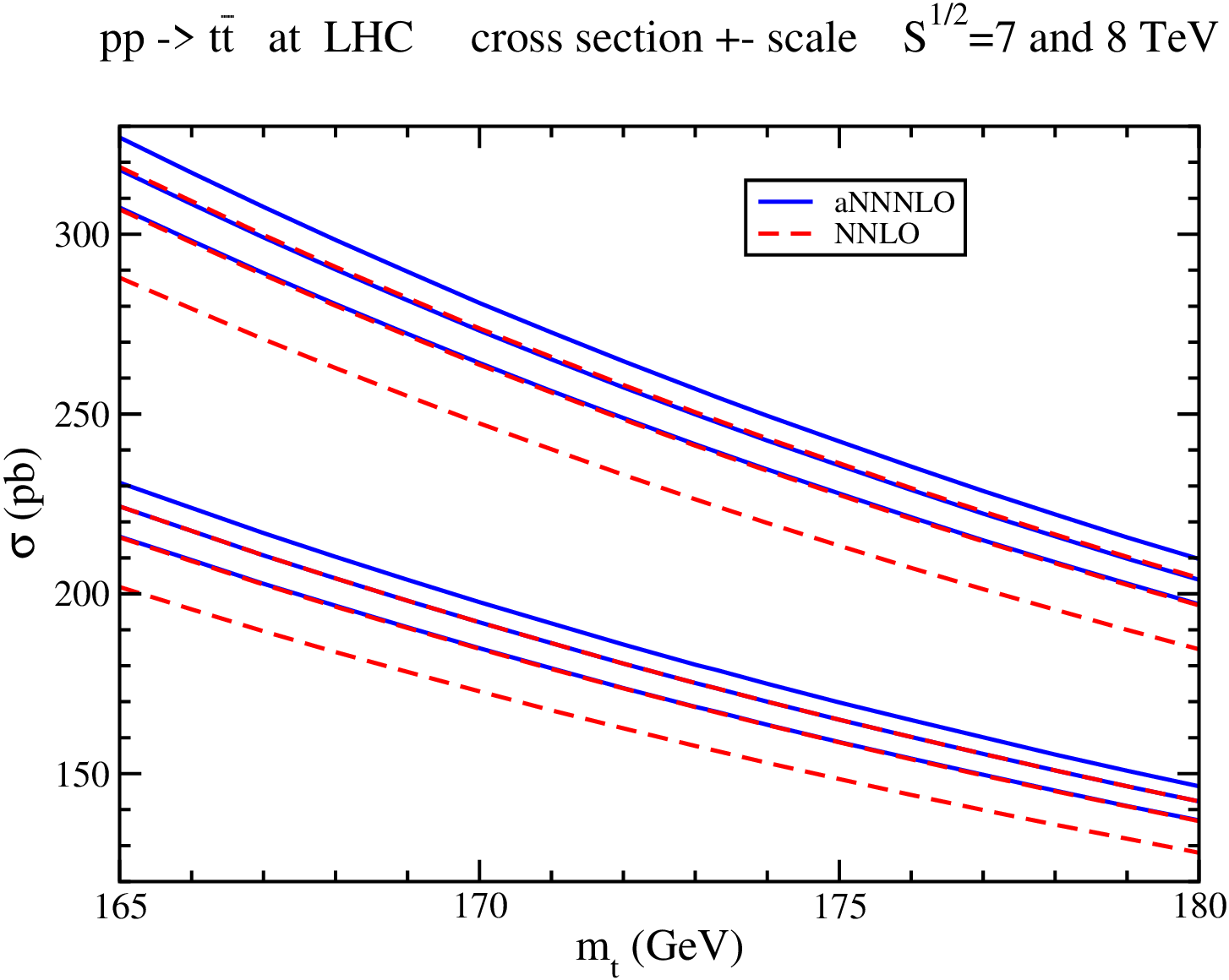}
\caption{The NNLO and aNNNLO total cross sections for $t{\bar t}$ production 
at the LHC with $\sqrt{S}=7$ TeV (lower lines) and 8 TeV (upper lines). 
The central lines at each order and each energy are with 
$\mu=m_t$, and the other lines display the upper and lower values from scale 
variation.}
\label{top7and8lhc}
\end{center}
\end{figure}

In Fig. \ref{top7and8lhc} we plot the NNLO and aNNNLO cross sections  
for top-antitop production at the LHC at 7 TeV and 8 TeV energies 
over a top quark mass range $165 \le m_t \le 180 $ GeV. 
The central results at each order and for each energy are with a factorization and renormalization scale $\mu=m_t$, and the scale variation is also displayed. We see that the upper NNLO curves are on top of the central aNNNLO curves 
at both energies; similarly the central NNLO curves are on top of the lower 
aNNNLO curves. The overall scale dependence is smaller at aNNNLO.

For a top quark mass of 173.3 GeV, the aNNNLO approximate cross section at 
7 TeV LHC energy is 
\beq
\sigma^{\rm aNNNLO}_{pp \rightarrow t{\bar t}}(m_t=173.3\, {\rm GeV}, \, \sqrt{S}=7\, {\rm TeV})
=174 {}^{+5}_{-7}  {}^{+9}_{-10} = 174^{+11}_{-12} \; {\rm pb}
\eeq
where the first uncertainty is from scale variation over $0.5 \le \mu/m_t \le 2$ 
and the second is from the MSTW2008 NNLO pdf errors at 90\% C.L.
The total uncertainty is +6.2\% $-6.8$\%. 
The aNNNLO enhancement over NNLO is 4.0\%.

At 8 TeV the corresponding result is 
\beq
\sigma^{\rm aNNNLO}_{pp \rightarrow t{\bar t}}(m_t=173.3\, {\rm GeV}, \, \sqrt{S}=8\, {\rm TeV})
=248 {}^{+7}_{-8}  {}^{+12}_{-13} = 248^{+14}_{-15} \; {\rm pb} \, .
\eeq
The total uncertainty is +5.7\% $-6.2$\%.
The aNNNLO enhancement over NNLO is 3.6\%.

\begin{figure}
\begin{center}
\includegraphics[width=11cm]{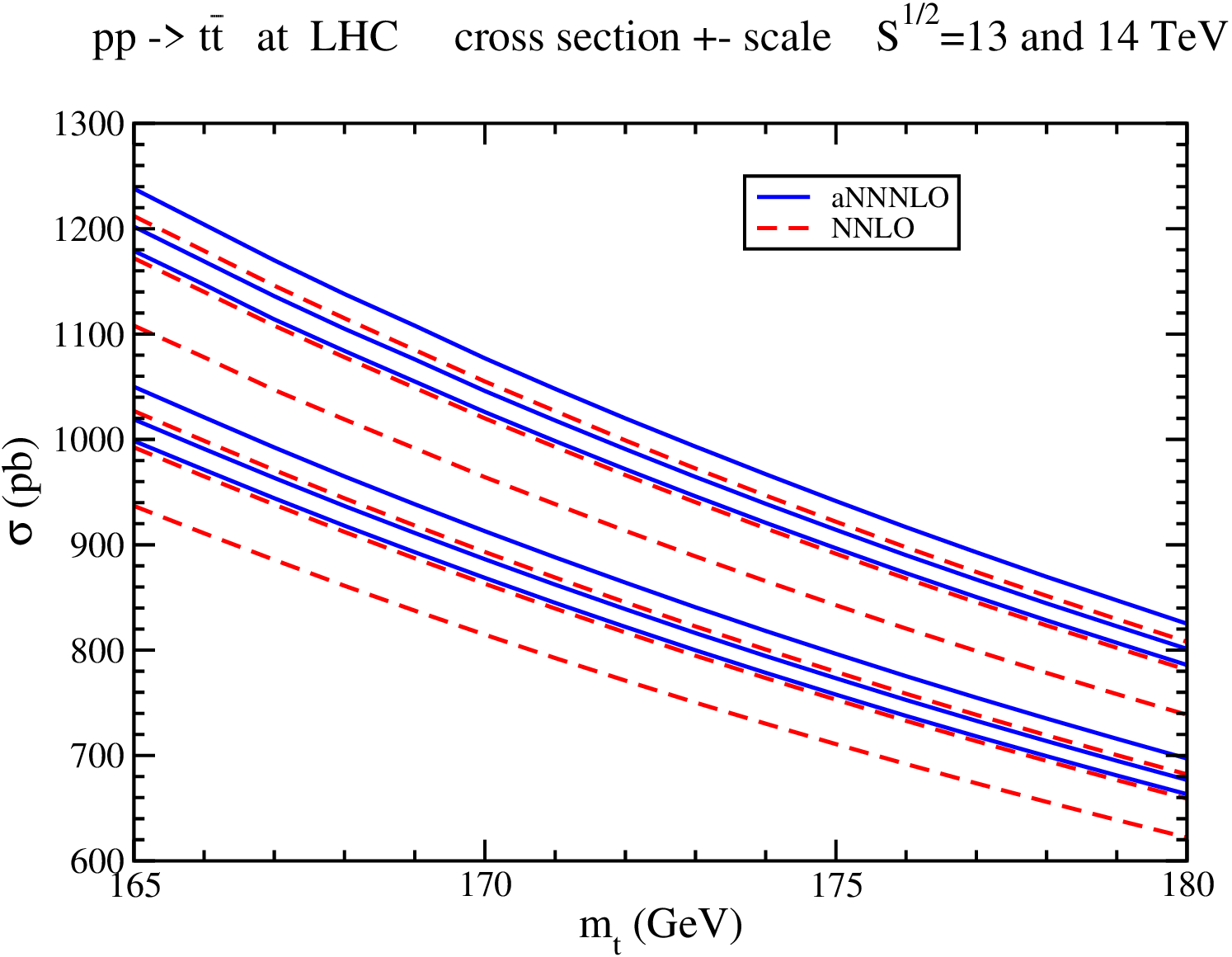}
\caption{The NNLO and aNNNLO total cross sections for $t{\bar t}$ production 
at the LHC with $\sqrt{S}=13$ TeV (lower lines) and 14 TeV (upper lines). 
The central lines at each order and each energy are with 
$\mu=m_t$, and the other lines display the upper and lower values from scale 
variation.}
\label{top13and14lhc}
\end{center}
\end{figure}

In Fig. \ref{top13and14lhc} we plot the NNLO and aNNNLO cross sections 
for top-antitop production at the LHC at 13 TeV and 14 TeV energies 
over a top quark mass range $165 \le m_t \le 180 $ GeV. 
Again the central results at each order and for each energy are with scales $\mu=m_t$, and the scale variation is also displayed. Again, the overall scale dependence is smaller at aNNNLO than at NNLO.

The approximate aNNNLO cross section at 13 TeV is 
\beq
\sigma^{\rm aNNNLO}_{pp \rightarrow t{\bar t}}(m_t=173.3\, {\rm GeV}, \sqrt{S}=13\, {\rm TeV})
=810 {}^{+24}_{-16}{}^{+30}_{-32}=810^{+38}_{-36} \; {\rm pb} \, . 
\eeq
The total uncertainty is +4.8\%  $-4.4$\% and
the aNNNLO enhancement over NNLO is 2.7\% at 13 TeV.

The approximate aNNNLO cross section at 14 TeV is
\beq
\sigma^{\rm aNNNLO}_{pp \rightarrow t{\bar t}}(m_t=173.3\, {\rm GeV}, \sqrt{S}=14\, {\rm TeV})
=957 {}^{+28}_{-19}{}^{+34}_{-36}=957^{+44}_{-41} \; {\rm pb} \, .
\eeq
The total uncertainty at 14 TeV is +4.6\% $-4.3$\% 
and the aNNNLO enhancement over NNLO is 2.6\%.

\begin{table}[htb]
\begin{center}
\begin{tabular}{|c|c|c|c|c|c|} \hline
\multicolumn{6}{|c|}{Fractional contributions to the perturbative series for the $t{\bar t}$ cross section} \\ \hline
corrections & Tevatron & LHC 7 TeV & LHC 8 TeV & LHC 13 TeV& LHC 14 TeV \\ \hline
$\sigma^{(1)}/\sigma^{(0)}$ & 0.236 & 0.470 & 0.476 & 0.493 & 0.496 \\ \hline 
$\sigma^{(2)}/\sigma^{(0)}$ & 0.106 & 0.178 & 0.177 & 0.172 & 0.170 \\ \hline 
$\sigma^{(3)}/\sigma^{(0)}$ & 0.068 & 0.066 & 0.059 & 0.045 & 0.043 \\ \hline 
\end{tabular}
\caption[]{The fractional contributions at higher orders relative to LO, 
as defined in Eq. (\ref{pertseries}), all calculated with the same 
NNLO pdf \cite{MSTW2008}, 
to the $t{\bar t}$ production 
cross section in $p{\bar p}$ collisions at the Tevatron 
with $\sqrt{S}=1.96$ TeV and in $pp$ collisions at the LHC with 
$\sqrt{S}=7$, 8, 13, and 14 TeV, with $\mu=m_t=173.3$ GeV.}
\label{table2}
\end{center}
\end{table}

It is interesting to discuss the convergence of the perturbative expansion.
We write the perturbative series for the cross section through aNNNLO as 
\beq
\sigma^{\rm aNNNLO}=\sigma^{(0)} \left[1+\frac{\sigma^{(1)}}{\sigma^{(0)}}
+\frac{\sigma^{(2)}}{\sigma^{(0)}}+\frac{\sigma^{(3)}}{\sigma^{(0)}}\right] 
\label{pertseries}
\eeq 
where $\sigma^{(0)}$ is the LO cross section, $\sigma^{(1)}$ denotes 
the complete NLO corrections, $\sigma^{(2)}$ denotes the complete NNLO 
corrections, and $\sigma^{(3)}$ denotes the NNNLO soft-gluon corrections.
The values for the fractions $\sigma^{(n)}/\sigma^{(0)}$, with $n=1,2,3$, 
appear in Table \ref{table2}.

The perturbative series converges slowly. At 14 TeV, $\sigma^{\rm aNNNLO}=1.709 \, \sigma^{(0)}$ as can be seen by summing the corresponding entries in Table \ref{table2} for Eq. (\ref{pertseries}); the NLO corrections are about half 
of the LO cross section; the NNLO corrections are about a third of the NLO 
corrections; and the aNNNLO corrections are about a quarter of the NNLO corrections. Thus the series $\sum_{n=1}^4 1/n!=1.708\cdots$ approximates the situation very well. If this trend continues in higher orders, which is of course not known but it is amusing to consider the possibility, then using the result  $\sum_{n=1}^{\infty} 1/n!=e-1=1.718\cdots$ one would expect less than 1\% of the LO cross section (or 0.5\% of the aNNNLO cross section) from  additional corrections beyond NNNLO. The above remarks describe the 13 TeV results very well too.

The convergence numbers for 7 and 8 TeV LHC energies are similar though in this 
case the aNNNLO corrections are about a third of the NNLO corrections, so the trend might me better described by the series $1+(1/2) \sum_{n=0}^2 1/3^n$. If this trend continues in higher orders, which again is not known, then using the result  $1+(1/2)\sum_{n=0}^{\infty} 1/3^n=7/4=1.75$ we again would not expect significant corrections beyond NNNLO.

At the Tevatron, however, the situation is quite different due to the fact that we are closer to threshold. Though the NLO corrections are only about a quarter 
of the LO cross section (and this is due to the fact that the $gg$ channel is numerically small in this case), the further effect of NNLO and aNNNLO corrections is significant and the convergence is slower. Again, this is due to the fact that we are closer to threshold and it is in agreement with the relatively large cross section measured at the Tevatron. In fact it is clear that as we progress in energy at the LHC from 7 through 14 TeV, and thus as we move away from threshold, the NNLO and aNNNLO corrections contribute a progressively smaller fraction to the overall cross section.

It is also important to consider the effects of the order of the pdf being used in the calculation. Since there are no NNNLO pdf available, the aNNNLO cross section that we find using NNLO pdf likely overestimates the result that we would get if we had used NNNLO pdf.
It is known that the effect is quite significant when going from NLO pdf to NNLO pdf. For example at 14 TeV LHC energy the NLO cross section using NLO pdf is 885 pb, but using NNLO pdf it is 838 pb. Although the effect from NNLO pdf to NNNLO pdf should be substantially smaller, it may be comparable to (and partially cancel out) the effect of additional corrections at NNNLO and beyond in the perturbative series. 

The above discussions indicate that our aNNNLO results are reliable and accurate, with any additional corrections due to higher orders and higher-order pdf being small and partly canceling out.

\begin{figure}
\begin{center}
\includegraphics[width=11cm]{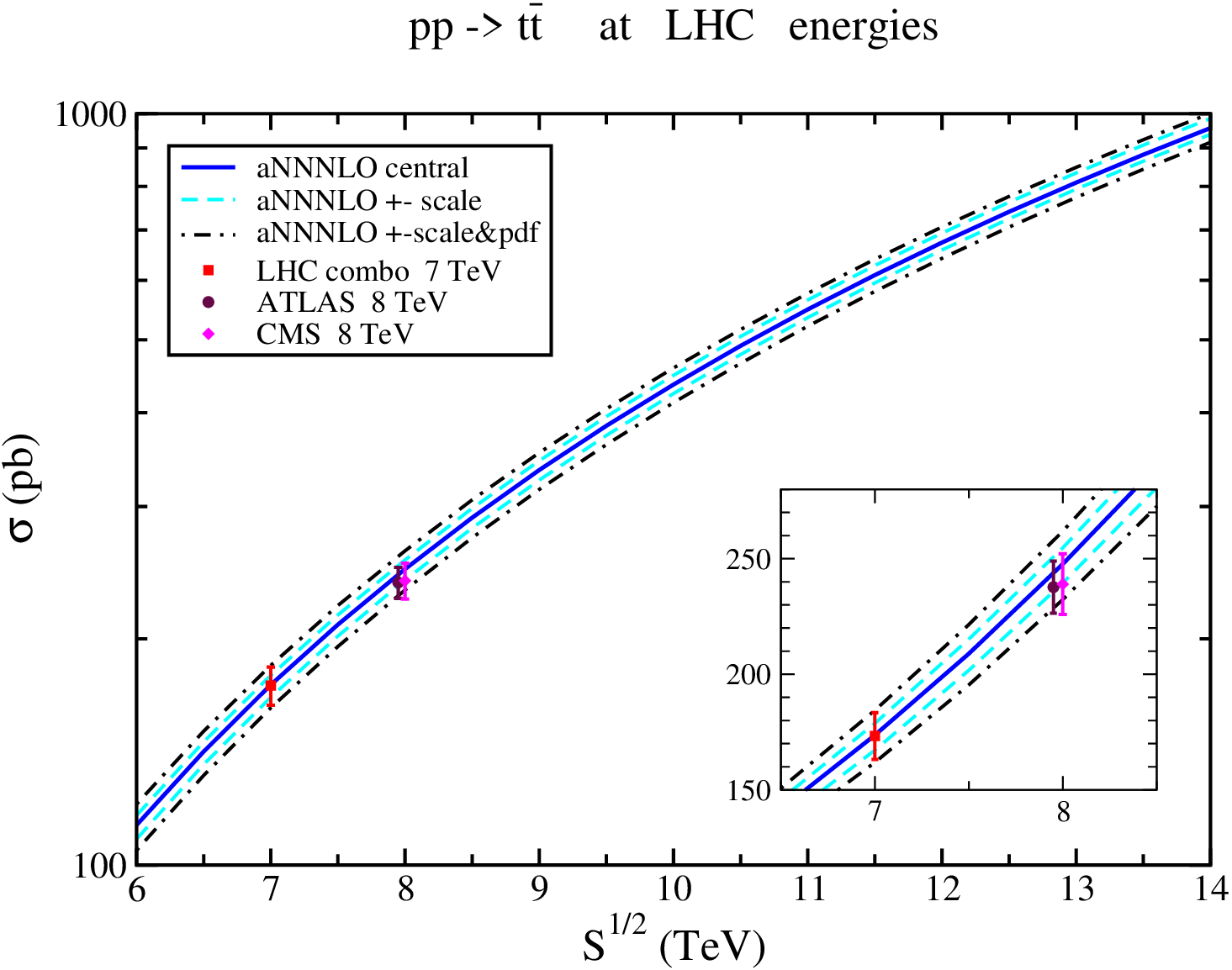}
\caption{The aNNNLO total cross section with scale and pdf uncertainties for 
$t{\bar t}$ production at the LHC as a function of collider energy. The inset 
plot highlights the comparison of theory with ATLAS and CMS data at 7 TeV \cite{LHC7} and 8 TeV \cite{ATLAS,CMS} LHC energies.}
\label{topSlhc}
\end{center}
\end{figure}

In Fig. \ref{topSlhc} we show the aNNNLO $t{\bar t}$ cross section as a function of LHC energy. In addition to the central result we show lines indicating 
the scale variation, as well as upper and lower lines indicating the range from the combined scale and pdf uncertainties. Very good agreement is found with recent ATLAS and CMS data at both 7 TeV \cite{LHC7} and 8 TeV \cite{ATLAS,CMS} energies. The experimental uncertainties are comparable to the theoretical ones. 

\begin{table}[htb]
\begin{center}
\begin{tabular}{|c|c|c|c|c|c|c|} \hline
\multicolumn{7}{|c|}{Fit parameters for the aNNNLO  $t{\bar t}$ cross section} \\ \hline
fit & parameters & Tevatron & LHC 7 TeV & LHC 8 TeV & LHC 13 TeV& LHC 14 TeV \\ \hline
\multirow{3}{*} & $\sigma(173.3)$ & 7.365 & 173.7 & 247.7 & 809.7 & 956.8 \\   
central & $c_1$ & 1.462 & 1.242 & 1.091 & 0.7087 & 0.6564 \\ 
& $c_2$ & 0.9459& 0.9175 & 0.7429 & 0.2165 & 0.1673 \\ \hline 
\multirow{3}{*}  & $\sigma(173.3)$ & 7.455 & 178.8 & 254.7 & 834.0 & 985.4 \\ 
+scale & $c_1$ & 1.469 & 1.249 & 1.099 & 0.7146 & 0.6630 \\
& $c_2$ & 0.9690 & 0.9350 & 0.7474 & 0.2264 & 0.1795 \\ \hline
\multirow{3}{*}  & $\sigma(173.3)$ & 7.091 & 167.2 & 239.5 & 793.5 & 938.5 \\ 
$-$scale & $c_1$ & 1.457 & 1.241 & 1.092 & 0.7085 & 0.6564 \\
& $c_2$ & 0.9268 & 0.9192 & 0.7377 & 0.2176 & 0.1673 \\ \hline
\multirow{3}{*}  & $\sigma(173.3)$ & 7.754 & 184.4 & 261.9 & 848.3 & 1001.1 \\ 
+scale \& pdf & $c_1$ & 1.421 & 1.220 & 1.070 & 0.6960 & 0.6451 \\
& $c_2$ & 0.8846 & 0.8859 & 0.7313 & 0.2095 & 0.1591 \\ \hline
\multirow{3}{*}  & $\sigma(173.3)$ & 6.975 & 161.9 & 232.3 & 773.7 & 915.9 \\ 
$-$scale \& pdf & $c_1$ & 1.481 & 1.267 & 1.1157 & 0.7277 & 0.6743 \\
& $c_2$ & 0.9774 & 0.9493 & 0.7760 & 0.2346 & 0.1813 \\ \hline
\end{tabular}
\caption[]{Values of the fit parameters in Eq. (\ref{paramfit}) for the aNNNLO $t{\bar t}$ production cross section in pb 
in $p{\bar p}$ collisions at the Tevatron with $\sqrt{S}=1.96$ TeV and   
in $pp$ collisions at the LHC with $\sqrt{S}=7$, 8, 13, and 14 TeV. 
The fits provide per mille or better accuracy for 130 GeV $\le m_t \le$ 210 GeV.}
\label{table3}
\end{center}
\end{table}

We can parameterize the aNNNLO cross sections as functions of the top-quark mass 
\beq
\sigma^{\rm aNNNLO}(m_t)=\sigma(173.3) \frac{173.3^4}{m_t^4} 
\left(1+c_1 \frac{(173.3-m_t)}{173.3}+c_2 \frac{(173.3-m_t)^2}{173.3^2}\right)
\label{paramfit}
\eeq
and we provide the results for the aNNNLO cross sections at $m_t=173.3$ GeV, denoted as $\sigma(173.3)$, and for the best fit values for the parameters $c_1$ and $c_2$ to four significant figures in Table \ref{table3}. 
The central results are with $\mu=m_t$; the results denoted as +scale and 
$-$scale are with scale variation between $m_t/2$ and $2m_t$; and the results 
denoted as ``+scale\&pdf'' and ``$-$scale\&pdf'' give the upper and lower values for the cross section, including scale variation and 90\% C.L. pdf uncertainties added in quadrature. These fits give the aNNNLO cross section with per mille or better accuracy for a large range of top-quark masses, 130 GeV $\le m_t \le$ 210 GeV. The numbers in Tables \ref{table1} and \ref{table2}, in the equations and discussions in the text, and in the figures can all be easily reproduced from these fits.

\mysection{Conclusions}

We have provided results for aNNNLO total cross sections for $t{\bar t}$ 
production at the LHC at 7, 8, 13, and 14 TeV energies and at the Tevatron 
at 1.96 TeV energy. The NNNLO soft-gluon corrections provide enhancements over the NNLO results which depend on the collider energy and vary from 5.1\% at the Tevatron 
to 3.6\% at the 8 TeV LHC and 2.6\% at the 14 TeV LHC.
Theoretical scale and pdf uncertainties have also been determined. The scale 
uncertainties are significantly reduced at aNNNLO relative to NNLO. The 
overall enhancements and uncertainty reductions provided by the aNNNLO 
corrections enable more accurate and precise theoretical predictions 
which are also in very good agreement with the latest results from the LHC 
and the Tevatron. 
Calculations of aNNNLO top-quark $p_T$ and rapidity distributions 
are also possible in our double-differential resummation formalism 
and will appear in the future.

\mysection*{Acknowledgements}
This material is based upon work supported by the National Science Foundation 
under Grant No. PHY 1212472.

\end{document}